\documentclass[aps,superscriptaddress,twocolumn]{revtex4-1}
\usepackage{epsfig}
\usepackage{amsmath,amssymb,amsthm}
\usepackage{graphicx}% Include figure files
\usepackage{psfrag}% use latex text  in ps figures
\usepackage{bm}% bold math  

\def\s2{\sigma^2}   
  
\begin{document}

\title{Revivals of Zitterbewegung of a bound localized Dirac particle}

\author{ E. Romera } 
\affiliation{Instituto Carlos I de F{\'\i}sica Te\'orica y 
Computacional, Universidad de Granada, Fuentenueva s/n, 18071 Granada, 
Spain} 
\affiliation{Departamento de F\'isica At\'omica, Molecular y Nuclear, 
Universidad de Granada, Fuentenueva s/n, 18071 Granada, Spain}

\date{\today}  
  
\begin{abstract}  
We show that a bound localized Dirac particle exhibits a revival of the
Zitterbewegung (ZB) oscillation amplitude. These revivals go  beyond the known quasiclassical
regenerations in which the ZB  oscillation amplitude is decreasing
from period to period. We study this phenomenon in a
Dirac-oscillator and show that it is  possible to set up wave-packets in which
there is a regeneration of the initial ZB  amplitude.

 \end{abstract}  
\maketitle  
  
\section{Introduction}  
In the context of relativistic quantum mechanics there is a surprising  phenomenon introduced by Schr\"odinger in 1930 as
Zitterbewegung (ZB) \cite{Schrodinger}. He showed that there is  a rapid ``trembling'' motion
of a Dirac particle around its otherwise rectilinear average trajectory, due to the interference between negative
and positive energy eigenvalues. There  has been a lot of theoretical studies
of ZB, but not a direct observation, due to the fact that the predicted frequency
and amplitude are impossible to measure  experimentally  at present. Lock showed that ZB
has a transient character for a free localized Dirac particle pointing out
that  ZB effect for a localized wave-packet in an external field depends on
the eigenvalues of the Hamiltonian \cite{Lock}. Nowadays, there is  an intense interest in ZB of
electrons in semiconductors  (see the review of Zawadzki and
Rusin \cite{Zawadzki} and references therein). Recently,  ZB has been studied in
graphene  \cite{Zawadzki,17,41,46,47,48,51,54,55} where it has been related to
electric conductivity. In particular, revivals and  ZB  were studied  in the
electric current in monolayer
graphene in a perpendicular magnetic field \cite{54}.
 In 2010,
Gerritsma et al. \cite{gerritsma}  simulated experimentally the
electron ZB by means of trapped ions and laser excitations adjusting
experimentally some parameters of the Dirac equation.

On the other hand, quantum revival  of wave-packets is an interference quantum phenomenon related to the
temporal evolution of wave-packets, relativistic and nonrelativistic. Quantum revivals have been
investigated theoretically  including in atomic, molecular and nonlinear
systems \cite{1,3,primenumber,rob,nes,nes1,strange} and observed experimentally in a lot of different
quantum systems, such as Rydberg atoms and molecules, and Bose-Einstein
condensates \cite{2,rob}. 

In what follow we show that there is a revival of ZB
oscillations  amplitude when a bound Dirac-electron is consider. We have chosen
a Dirac oscillator to analyze this  behavior because it is exactly soluble
and it is a model that has applications in several branches of physics (see
\cite{Mandal} and references therein). In this work it is demonstrated that
besides  ZB and quasiclassical oscillations studied previously by other
authors \cite{Bermudez}, there exists a revival or regeneration of the ZB oscillations amplitude. 

To describe
quantum revivals, let us consider an
initial wave-packet that is a superposition of eigenstates localized around  
some energy level $E_{n_0}$, it is appropriate to expand the energy  around $n_0$ if $|n-n_0|/n_0<<1$,
\begin{equation}
E_n\approx E_{n_0}
+E^{\prime}_{n_0}(n-n_0)+\frac{E^{\prime\prime}_{n_0}}{2}(n-n_0)^2 + \cdots
\label{series}
\end{equation}
and each term in the series defines an important time scale, 
$
T_{CL}=\frac{2\pi\hbar}{|E^{\prime}(n_0)|}$,
$T_{R}=\frac{2\pi\hbar}{|E^{\prime\prime}(n_0)|/2}$
where $T_{CL}$ is  associated with the classical periodic motion of the
wave-packet and $T_{R}$ is the revival time (the validity of this expansion has been demonstrated in
\cite{3,6n,15n}). The wave-packet initially evolves
quasiclassically with period $T_{CL}$, then spreads and collapses, but at later
times, around $T_{R}$ the wave-packet regenerates and  reaches approximately  its initial
shape. For times that are rational fractions of $T_R$ wave-packets split in
clones of themselves \cite{4,rob}. After the revival time a new cycle starts
with quasiclassically behavior, collapses, fractional revivals and
revivals. Revivals are usually analyzed using the autocorrelation function
$A(t)$, which is the overlap between the initial and the time-evolving
wave-packet. An alternative approach in terms  on uncertainty entropic
relations was  proposed \cite{nosotros}.

\section{Revivals of Zitterbewegung in a Dirac oscillator}

\begin{figure}
\includegraphics[width=9cm]{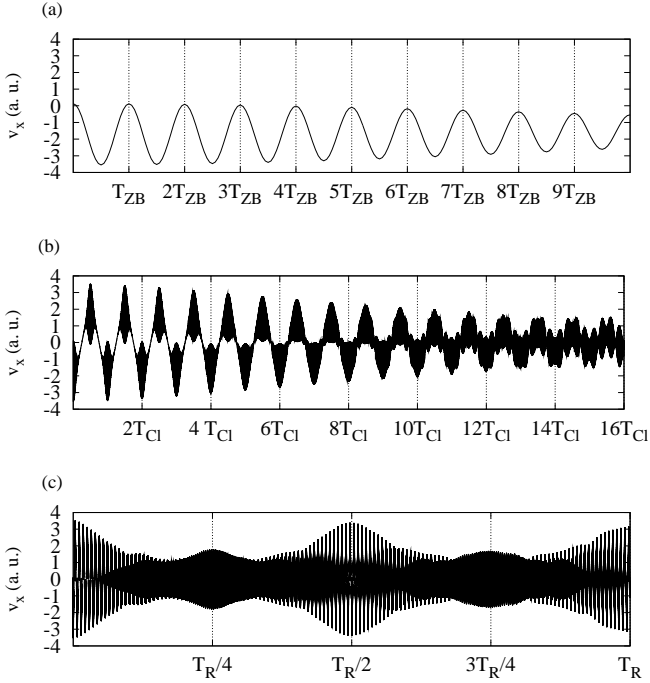}
\caption{ Time dependence of $\langle v_x\rangle$ for the initial wave-packet
  with $\sigma=3$, $n_0=30$, and oscillator frequency $\omega=10^3$,
  for which $T_{ZB}= 6.15\times 10^{-5}$ , $T_{CL}=8.54\times 10^{-3}$ , and $T_{R}=1.19$
  (all in a. u.). The vertical dotted lines stand for (panel a) $T_{ZB}$
  periods (panel b) $T_{CL}$ periods, and (panel c) $T_R$ periods.}
\label{figzb}
\end{figure}
\begin{figure}
\includegraphics[width=9cm]{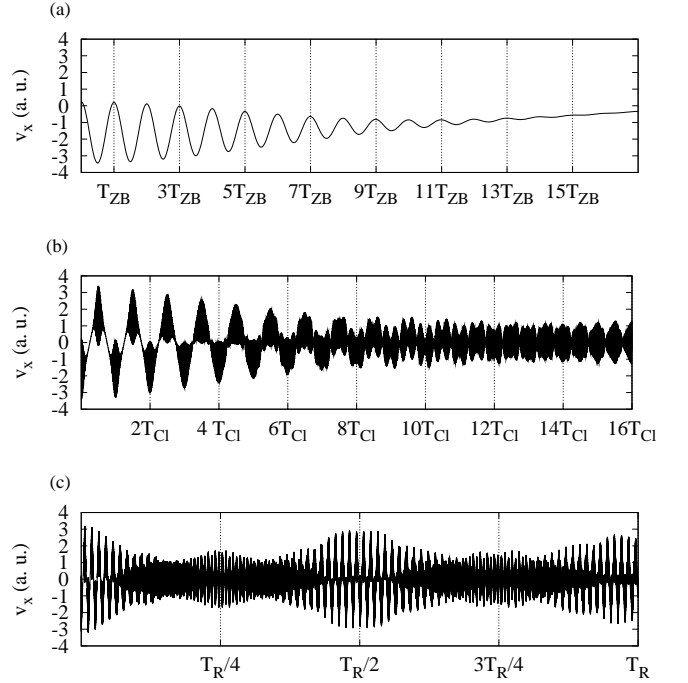}
\caption{ Time dependence of $\langle v_x\rangle$ for the initial wave-packet
  with $\sigma=3$, $n_0=15$, and oscillator frequency $\omega=10^3$,
  for which $T_{ZB}= 8.16\times 10^{-5}$ , $T_{CL}=6.4\times 10^{-3}$ , and $T_{R}=0.5$
  (all in a. u.). The vertical dotted lines stand for (panel a) $T_{ZB}$
  periods (panel b) $T_{CL}$ periods, and (panel c) $T_R$ periods.}
\label{figzb2}
\end{figure}
An appropriate system to discuss revivals  of ZB for bounded
states is a (2+1)-Dirac oscillator, due to the fact that it is exactly soluble
and allows us to study this phenomenon in a simple system. So, we
shall consider the Hamiltonian for a Dirac oscillator \cite{Moshinsky} with
frequency $\omega$
\begin{equation}
H = c {\bf \alpha}\cdot ({\bf p} - i m\omega\beta{\bf r}) + \beta m c^2
\label{diracequation}
\end{equation}
where $m$ is the rest mass of the Dirac particle (p. e. an electron), ${\bf \alpha}$ and $\beta$ are the
Dirac matrices, and $c$ the speed of light.
We shall introduce the complex coordinate as in \cite{Mandal}  $z = x + i y$ and
using the usual creation and annihilation operators  notation in terms of $z$
and $\bar{z}$
\[
a=\frac{1}{\sqrt{m\omega \hbar}}p_{\bar{z}} -
  \frac{i}{2}\sqrt{\frac{m\omega}{\hbar}}z 
\]
\[
a^{+}=\frac{1}{\sqrt{m\omega \hbar}}p_{z} +
  \frac{i}{2}\sqrt{\frac{m\omega}{\hbar}}\bar{z}
\]
the Hamiltonian reads
\begin{equation}
H= \begin{pmatrix} mc^2& 2c\sqrt{m\omega\hbar}a^{+}\\ 2c\sqrt{m\omega\hbar}a  & -mc^2 \end{pmatrix}.
\label{hamiltoniano}
\end{equation}
It is not difficult to show that the energy eigenfunctions are given by
\begin{equation}
|\phi_n^{\pm}\rangle=\begin{pmatrix} \pm\sqrt{\frac{1}{2}\pm \xi_n}|n\rangle \\
  \mp\sqrt{\frac{1}{2}\mp \xi_n}|n-1\rangle  \end{pmatrix},
\end{equation}
with \begin{equation}\xi_n=\frac{1}{2\sqrt{1+\frac{4\hbar\omega n}{mc^2}}}\end{equation} and with
$n=0,1,...$ and the energy spectrum is, in turn, \begin{equation} E^{\pm}_n=\pm mc^2\sqrt{1+\frac{4\hbar\omega n}{mc^2}}.\end{equation} 

 We shall construct  a superposition state of  two wave-packets as the
initial particle wave-packet
\begin{equation}
|\Psi_0\rangle = \frac{1}{\sqrt{2}}(|\Psi_-\rangle + |\Psi_+\rangle)
\label{initial}
\end{equation}
 where the above  wave packets are defined  as the linear combination

\begin{equation}
|\Psi_{+}\rangle= \sum_{n} c^{+}_{n} |\phi^{+}_{n}\rangle\quad 
\mbox{  and  } \quad |\Psi_{-}\rangle= \sum_{n} c^{-}_n |\phi^{-}_n\rangle,
\label{ini1}
\end{equation}
each of them  centered around a given eigenvalue $E^{+}_{n_0}$ and $E^-_{n_0}$,
respectively, with coefficients Gaussianly distributed ($c_n^{+}=c_n^{-}=c_n$)
as
\begin{figure}
\includegraphics[width=9cm]{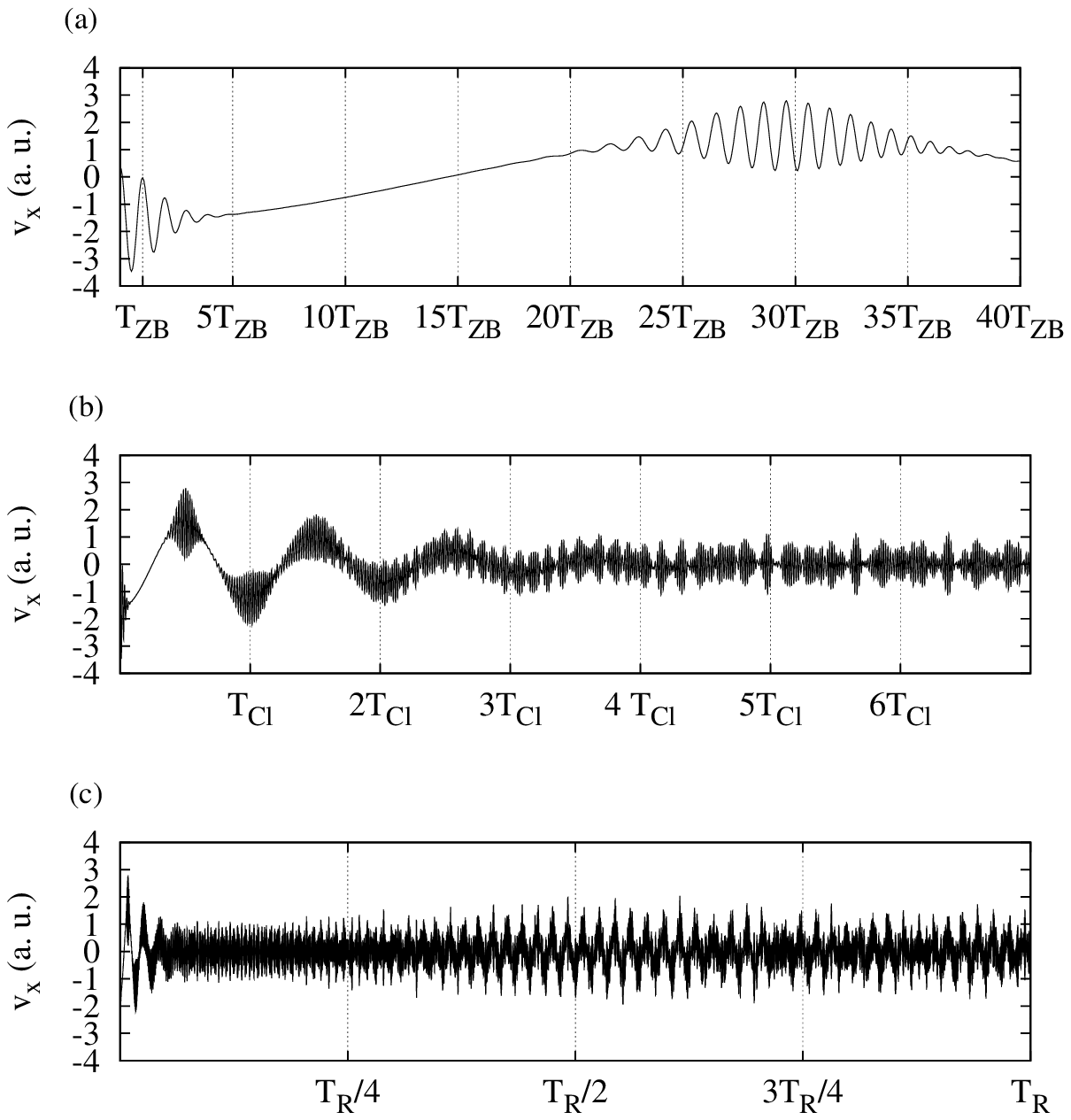}
\caption{ Time dependence of $\langle v_x\rangle$ for the initial wave-packet
  with $\sigma=20$, $n_0=10$, and oscillator frequency $\omega=10^3$,
  for which $T_{ZB}= 9.46\times 10^{-5}$ , $T_{CL}=5.55\times 10^{-3}$ , and $T_{R}=0.33$
  (all in a. u.). The vertical dotted lines stand for (panel a) $T_{ZB}$
  periods (panel b) $T_{CL}$ periods, and (panel c) $T_R$ periods.}
\label{figzb3}
\end{figure}
\begin{figure}
\includegraphics[width=9.cm]{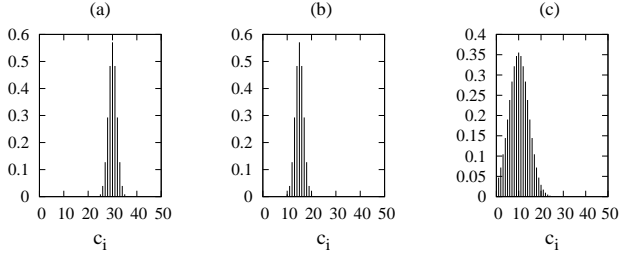}
\caption{Coefficients (Eq. \ref{coe1}) for (a) $n_0=30$, $\sigma=3$, (b) $n_0=15$,
  $\sigma=3$, and (c) $n_0=10$, $\sigma=20$.}
\label{coeficientes}
\end{figure}
\begin{equation}
c_n=\sqrt{\frac{1}{\pi\sqrt{\sigma}}} e^{-(n-n_0)^2/2\sigma}.
\label{coe1}
\end{equation}
We can write the  temporal evolution of the initial wave-packet as
 \begin{equation}|\Psi_0(t)\rangle=\frac{1}{\sqrt{2}}\sum_{n}( c^{+}_{n} |\phi^{+}_{n}\rangle
 e^{iE^+_nt/\hbar}
 + c^{-}_n |\phi^{-}_n\rangle e^{iE^-_nt/\hbar} )
\end{equation} 
taking into account that
\begin{equation}
|\Psi_{\pm}(t)\rangle=\sum_{n}( c^{\pm}_{n} |\phi^{\pm}_{n}\rangle
 e^{iE^{\pm}_nt/\hbar}.
\label{aa}
\end{equation}

The series expansion (\ref{series}) should  be interpreted in the context of
the  temporal
evolution of the wavepacket. If we replace the value $E_n$ by the expansion
(\ref{series}) in Eq. (\ref{aa}) we can see that each term in the exponential
(except the first) defines an important characteristic time scale. The first
term  is unimportant because it is an overall phase. The following two  terms define the
classical periodicity,  and  the revival time \cite{3,6n,15n}.

Therefore the corresponding classical period and the revival time for  $|\Psi\rangle$ yield straightforwardly
\begin{equation}
T_{CL}=\frac{\pi}{\omega}\sqrt{1+\frac{4\hbar\omega}{mc^2} n_0}\label{CL}\end{equation} and

\begin{equation}
T_{R}=\frac{\pi m c^2}{\hbar\omega^2}(1+\frac{4\hbar\omega}{ mc^2} n_0)^{3/2}
\label{R}.
\end{equation}  To calculate the
period of ZB we shall determine the temporal evolution of the   $x$ and $y$ components of the
velocity  which are given by $\langle v_j\rangle=\langle i[H,r_j]/\hbar\rangle$, ($j=x,y$), where
$\sigma_x$ and $\sigma_y$ are the Pauli matrices. For the  wave-packet
$|\Psi_0\rangle$, after some algebra and taking into account that $|n\rangle$ is an
orthonormal set,  the temporal evolution for the velocities is given by:
\begin{equation}
\begin{array}{l}
\langle v_x\rangle= 2\sum_{n=0}^{\infty}c_n c_{n+1}\left(\eta_n \cos((E_n+E_{n+1})t/\hbar) \right.\\\left.-\nu_n
  \cos((E_n-E_{n+1})t/\hbar)\right)\\
\langle v_y\rangle=0
\end{array}
\label{velocidad}
\end{equation}
where  $\eta_n=\gamma_n\gamma_{n+1}+\delta_n
  \delta_{n+1}$ and $\nu_n=\gamma_n\delta_{n+1}+\gamma_n \delta_{n+1}$
$\gamma_n=\sqrt{\frac{1}{2}+\xi_n}$ and $\delta_n=\sqrt{\frac{1}{2}-\xi_n}$ with
$n=0,1,2,...$. Several types of oscillatory motion emerge for the velocity evolution. The first term in the $v_x$ temporal evolution is weighted by 
 $\cos((E_n+E_{n+1})t/\hbar)$ which  is the responsible of the
ZB oscillatory motion. We estimate the ZB period  using equation
(\ref{series}) , which enables us to write $E_n+E_{n+1}\approx 2 E_{n_0}$  \cite{54} and
then 
\begin{equation}
T_{ZB}=\pi\hbar/|E_{n_0}|=\frac{\pi\hbar}{mc^2\sqrt{1+ \frac{4\hbar\omega}{m
      c^2}n}}.
\label{ZB}
\end{equation}
 The second term  in the $v_x$ temporal evolution is
weighted by  $ \cos((E_n-E_{n+1})t/\hbar)$ which lets us to extract different
periodicities in the velocity temporal evolution. Using Eq. (\ref{series})
again, we obtain other oscillatory scales $E_n-E_{n+1}\approx
E^{\prime}_{n_0}(n-n_0)+ E^{\prime\prime }_{n_0}(n-n_0)^2+...$, which are
given by $T_{CL}$ and $T_R$.

\begin{figure}
\includegraphics[width=9cm]{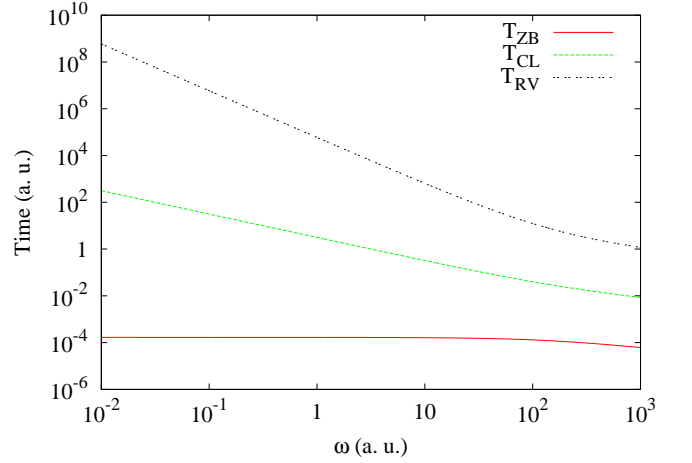}
\caption{ (Color online) Temporal scales $T_{ZB}$, $T_{CL}$ and $T_{R}$ vs $\omega$ (all in
  a. u.) for $n_0=30$.}
\label{tiempos}
\end{figure}

The velocity  behavior is clearly illustrated in figure \ref{figzb}. The value
$\langle v_x\rangle$ is numerically computed  as a function of time  for the
temporal evolution of the initial wave
packet $|\Psi_0\rangle$
 with $n_0=30$ and $\sigma=3.0$ and for an oscillator  frequency $\omega=10^3$
 a. u.. (Throughout the results are
generated in   atomic units $m=\hbar=e=1$). 
We have constructed the initial wave packet (see Eqs. \ref{initial},
\ref{ini1} and \ref{coe1} ) with the levels population given in
Fig. \ref{coeficientes}(a).
 We observe
 in panel (a) that 
 there is an oscillatory
 behavior for  $T_{ZB}$-time-scale. For  greater time-scales we can see in panel (b) that  
 quasiclassical oscillations appear enveloping the ZB oscillations whose
 amplitude is decreasing from period to period. This
 behavior was previously observed in \cite{Bermudez}.
 Finally in
 panel (c) we can clearly see  a new time-scale oscillation $T_R$, which is
 enveloping the previous oscillations, and it is apparent that for $t= m
 T_R/2$ (for $m=1,2$,...)  there is a revival
 of the ZB oscillation amplitude (Fig. 1 panel (c))  and the quasiclassical
 oscillations.
 
Let us illustrate this phenomenon with another example. In Fig. \ref{figzb2} we have considered an initial localized wave packet with a different
 value of the  parameter $n_0=15$ (That is, with the levels population given in
Fig. \ref{coeficientes}(b)). Then
    $T_{CL}$ and $T_{R}$ are smaller as $\omega$ is
 smaller
 and we observe that $T_{ZB}$ is somewhat
 lower than in the above case,  as we expected from equations
(\ref{CL}), (\ref{R}) and (\ref{ZB})
 respectively. Again we can observe a revival of the ZB amplitude (Fig. 2 panel (c)).

Moreover we have to stress that the appearance of  revivals of ZB oscillation
amplitude depends on the shape of the  initial wave packet, i. e.,  we
have to work with a localized wave-packet. If we  consider a broader   wave
packet 
around lower energies $\pm E_{n_0}$, with $n_0=10$ and $\sigma=20$ (Fig.
\ref{figzb3}) (see level population in Fig. \ref{coeficientes}(c)), we observe
in panel (a) that there is an oscillatory behavior similar to Fig. 1 and 2 for
the first quasiclassical periods where the ZB oscillation amplitude is
greater when the $|\langle v_x\rangle|$ is greater. But next, we can observe a
quasiclassical modulation that is disappearing in three classical periods
(Fig. 3 (b)) and we will have ZB but there is no regeneration of the initial
ZB amplitude 
($\approx 3.8$ a. u.). For much longer times
the revivals or quasiclassical behavior never appears (we have checked it from 0 to
10 $T_{R}$).

Finally, in Fig. \ref{tiempos}  we have studied the periods in terms of the parameter
omega. We can see that $T_{R}>T_{CL}>T_{ZB}$ and
 $T_{ZB}$ is almost constant for all $\omega$. $T_{CL}$ and $T_{R}$  increase when $\omega$
  decreases. In addition, when $\omega$ is smaller the temporal scales move away form each
 other quickly.  In fact the revival of the ZB amplitude  appears later.
 The revival of the ZB amplitude will disappear when $\omega=0$ (which
 corresponds with the Lock result \cite{Lock}). Note that in this limit case the ZB
 would be  approximately  $10^{-4}$ a. u..These results are an extension  for a bound Dirac particle  of the results   found for massless quasiparticles in graphene in a
perpendicular magnetic field \cite{54}.

 It should be noticed that the existence of  revivals  of the wave-packet, 
 and consequently  of the same initial quasiclassical behavior of $\langle
 v_x\rangle$ and  ZB oscillation amplitude,  is due to:
(i) the way in which we have constructed it   as a
superposition of two wave-packets localized around two given eigenvalues
$E_{n_0}^+$ and $-E_{n_0}^-$, and (ii)  the fact that the  Dirac oscillator has
electron-hole symmetry  ($E_n^+=-E_n^-$) which is an essential property to obtain Eqs. (14) and (15). 

Furthermore, a natural generalization of this result could be done   as follows. If we consider a bound Dirac particle with a non linear spectrum
$E_n^{\pm}$ in $n$ and with electron-hole symmetry, we  expect that the
localized Dirac particle  exhibits a revival of the ZB oscillation
amplitude. Although it is an  open problem to prove this assertion,   it could be
justified since one can always  consider an initial
wavepacket as a superposition of two localized wavepackets,  whith
the coefficients  centered around a mean value $n_0$ with $|n-n_0|<<n_0$
and  obtain an analogous behavior to Eq. (14)  for the  temporal
evolution  of the velocities. We have to remark  that the condition $E_n^+=-E_n^-$ is an
essential point to have  definite and visible temporal scales.
If  the initial wave-packet is  more localized and the $n_0$ value
is higher,   the revival of the ZB oscillation amplitude will be more
sharp due to the fact that the regeneration will be more accurate because the
Taylor expansion is more accurate too.

\section{Conclusions}

Summing up, we have studied the wave-packet dynamic for a Dirac oscillator
demonstrating that for some particular election of the initial wave-packet there is
a regeneration or revival of the ZB oscillation amplitude apart
from the quasiclassical  modulation of ZB in which the oscillation amplitude
is decreasing. These revivals appear associated with a nonlinearity in the
 relativistic eigenvalue spectrum.  When
the frequency of oscillation  is smaller the
regeneration appears at longer times. In the
limit of frequency zero, that is for a free Dirac particle, the regenerations
disappear, due to the fact that in the case of the  free Dirac particle the 
spectrum is continuous rather than discreet. 
We conjecture
  that this result may appear in any  bound Dirac particle with electron-hole symmetry.

\section{acknowledgments}

This work was supported by projects PYR-2010-24,  FIS2008-01143 and  FQM-165/0207.

\end{document}